\documentclass{article}

\usepackage[preprint]{neurips_2023_gaied}

\usepackage[utf8]{inputenc} 
\usepackage[T1]{fontenc}    
\usepackage{hyperref}       
\usepackage{url}            
\usepackage{booktabs}       
\usepackage{amsfonts}       
\usepackage{nicefrac}       
\usepackage{microtype}      
\usepackage{xcolor}         
\usepackage{todonotes}
\usepackage{amsmath}
\usepackage{pdfpages}
\usepackage{comment}
\usepackage{caption}

\usepackage{caption}
\usepackage{subcaption}
\usepackage{listings}

\setcitestyle{numbers}

\usepackage{tikz}
\newcommand*\circled[1]{\tikz[baseline=(char.base)]{
            \node[shape=circle,draw,inner sep=2pt] (char) {#1};}}
\newcommand*\circleddotted[1]{\tikz[baseline=(char.base)]{
            \node[shape=circle,draw,dotted,inner sep=2pt] (char) {#1};}}
            
\title{Benchmarking Educational Program Repair}

\author{%
  Charles Koutcheme \\
  Aalto University\\
  Espoo, Finland \\
  \texttt{charles.koutcheme@aalto.fi} \\
  \And
  Nicola Dainese \\
  Aalto University\\
  Espoo, Finland \\
  \texttt{nicola.dainese@aalto.fi}\\
  \And
  Sami Sarsa \\
  Aalto University\\
  Espoo, Finland \\
  \texttt{sami.sarsa@aalto.fi}
  \And
  Juho Leinonen \\ 
  University of Auckland\\
  Auckland, New Zealand \\
  \texttt{juho.leinonen@auckland.ac.nz} \\
  \And
  Arto Hellas \\ 
  Aalto University\\
  Espoo, Finland \\
  \texttt{arto.hellas@aalto.fi} \\
  \And
  Paul Denny \\ 
  University of Auckland\\
  Auckland, New Zealand \\
  \texttt{paul@cs.auckland.ac.nz}}
\begin{document}

\maketitle

\begin{abstract}
The emergence of large language models (LLMs) has sparked enormous interest due to their potential application across a range of educational tasks.  For example, recent work in programming education has used LLMs to generate learning resources, improve error messages, and provide feedback on code. However, one factor that limits progress within the field is that much of the research uses bespoke datasets and different evaluation metrics, making direct comparisons between results unreliable. Thus, there is a pressing need for standardization and benchmarks that facilitate the equitable comparison of competing approaches. One task where LLMs show great promise is program repair, which can be used to provide debugging support and next-step hints to students. In this article, we propose a novel educational program repair benchmark. We curate two high-quality publicly available programming datasets, present a unified evaluation procedure introducing a novel evaluation metric rouge@k for approximating the quality of repairs, and evaluate a set of five recent models to establish baseline performance.
\end{abstract}

\section{Introduction} 

In education, the emergence of large language models (LLMs) has led to a plethora of research evaluating their performance on educationally relevant tasks~\cite{kasneci2023chatgpt, denny2023computing}.  LLMs have been explored for educational tasks such as providing automatic feedback to students~\cite{pankiewicz2023large,kiesler2023exploring,hellas2023exploring}, generating novel exercises~\cite{sarsa2022automatic}, and producing novice-friendly explanations of source code~\cite{macneil2023experiences,leinonen2023comparing}.
Within computing education, one long-standing challenge has been generating automated feedback on students' programs~\cite{keuning2018systematic}. This has traditionally been achieved using unit-test-based automated assessment systems ~\cite{paiva2022automated,ihantola2010review}, 
or intelligent tutoring systems that provide hints on the next steps that students should take~\cite{gross2014example,rivers2017data,price2017evaluation}. 
A key parallel stream to this research is `automated program repair', which employs rule-based and machine-learning-based strategies for automatically fixing bugs in code~\cite{prenner2021automatic,koutcheme_automated_2023,joshi2022repair,ahmed2022synfix,phung2023generating}. 
Drawing from this extensive body of prior research, we recognize automated program repair as a crucial intermediary step toward enhancing the accuracy and reliability of feedback for novice programmers. Currently, substantial interest lies in harnessing LLMs for generating feedback based on program repairs \cite{zhang2022repairing,prenner2021automatic,koutcheme_automated_2023}. 

In general, much of the prior work exploring LLMs in education has used bespoke datasets due to the lack of widely used benchmarks.  This makes it difficult to meaningfully compare results and identify effective techniques. Benchmarks are an essential tool in computer science and software engineering, including machine learning. They provide a standardized method for comparing various techniques and for monitoring progress within an area over time, driving innovation and improvement.  Datasets in particular have served as benchmarks for comparing algorithms in a reproducible way which is the bedrock of modern science.  For example, within the field of machine learning standard benchmarks such as ImageNet~\cite{deng2009imagenet}, MNIST~\cite{lecun2010mnist} and HumanEval~\cite{chen2021evaluating} have been crucial for identifying breakthroughs in algorithms for image recognition and program synthesis. 


In this paper, we propose a framework for evaluating program repair techniques using LLMs. By providing a benchmark for this task, we hope to facilitate future research by allowing easy replication of results and evaluation of new models using the same data that older models have been evaluated with. 
In particular, this work makes the following contributions: we (1) formalize the task of educational program repair, (2) describe an evaluation procedure including a novel metric for the quality of program repairs called rouge@k, (3) curate two high-quality and publicly available datasets suitable for serving as a benchmark to evaluate program repairs, and (4) report the performance of several recent decoder-only transformer models on these datasets.

\section{Related Work}
 
\paragraph{Benchmarking program synthesis.} 
In assessing the capabilities of code language models across various contexts, programming benchmarks have become a common tool~\cite{chen2021evaluating,austin2021program,hendrycks2021measuring,lai2022ds1000}. 
However, when it comes to program synthesis, few of these benchmarks offer a large number of problems that are suitable for introductory programming courses~\cite{austin2021program,hendrycks2021measuring}. While code language models have traditionally been used in industry and research, the emergence of highly proficient easily accessible language models such as ChatGPT has sparked interest in Computing Education Research (CER). In line with program synthesis principles, extensive research in CER has explored these models' abilities to solve real-world introductory programming problems under authentic course conditions, moving beyond artificial examples found in standard code benchmarks~\cite{ansley2022_robots_are_coming,finnieansley2023myai,denny2023conversing_with_pilot,wermelinger2023using,savelka2023generative}. We notice a similar transition happening with program repair. 

\paragraph{Benchmarking program repair.}
The task of program repair is often approached as a synthesis problem. In our study, we draw inspiration from this paradigm to design components of our evaluation procedure. Similarly to program synthesis, there exist program refinement benchmarks \cite{lu2021codexglue,quixbugs} and particularly close to our work is a recent extension \cite{muennighoff2023octopack} to the human evaluation benchmark \cite{chen2021evaluating} that introduces bugs in code for LLMs to rectify. Although the bugs in all these datasets are relevant to student programming, they do not fully encapsulate the complexities of errors encountered in student-written code, which can present a wide array of issues extending beyond mere bugs. Consequently, while such refinement benchmarks assume the presence of ground truth annotations, our evaluation procedure operates without assuming the existence of a single ground truth repair, and, in fact, did not necessitate manual annotations.


Our work also builds upon recent efforts utilizing large language models to provide feedback and repair programs for students. Notably, LLMs have shown great promise in fixing syntax errors \cite{joshi2022repair,ahmed2022synfix,phung2023generating,leinonen2023using} and rectifying bugs in student programs \cite{zhang2022repairing}, and preliminary work exists on using them to automatically generate feedback~\cite{hellas2023exploring,kiesler2023exploring,pankiewicz2023large,liffiton2023codehelp,wu_2021_proto}. While most of these endeavours leverage closed-source systems such as OpenAI Codex, some explore the use of open-language models \cite{koutcheme_training_2023,koutcheme_automated_2023}.
Many of the methods we introduce align with these previous efforts. In particular, our proposed evaluation procedure aims to encompass the different scenarios explored in this prior research while leveraging recent insights into evaluation methodologies \cite{koutcheme_2023_evaluating}. 
Although some prior work has integrated human judgment into the evaluation~\cite{phung2023generative,leinonen2023using}, we focus on a fully automated evaluation approach which is more scalable, primarily relying on program correctness and NLP scoring metrics as proxies for quality. 
In addition to presenting the evaluation procedure, we also identify two publicly available datasets that we curate in such a way that they are suitable as benchmarks for evaluating program repair.



\section{Educational Program Repair}
\begin{figure}[htpb]
\centering
\includegraphics[width=\textwidth]{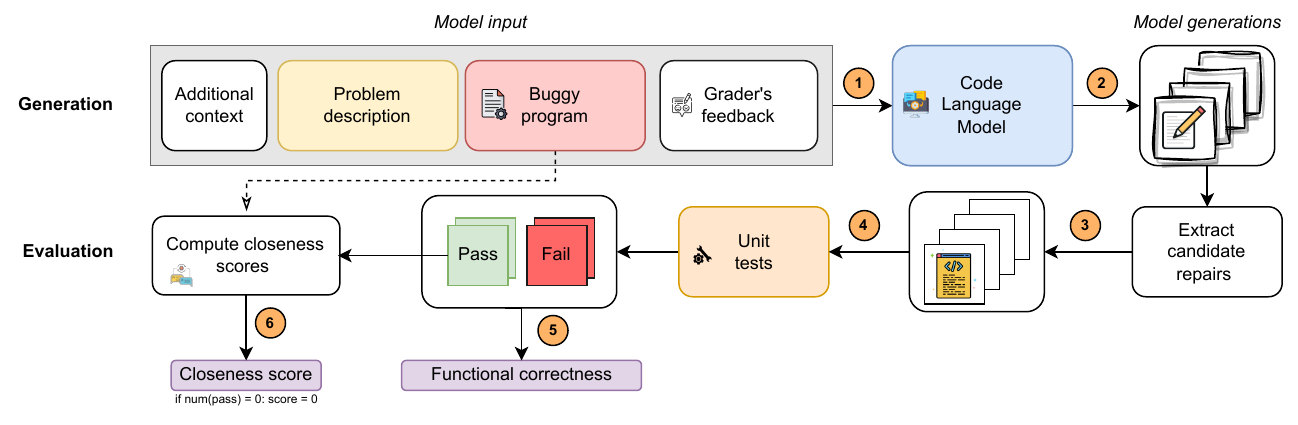}
\caption{\textbf{Overview of the program repair task and its evaluation procedure.} (1) For each incorrect program, we pass to a language model a prompt which contains the programming task, and the incorrect student code. 
In addition, the prompt may contain more contextual information, e.g. grader's feedback.
(2) The model generates multiple feedbacks based on the instructions; each feedback must contain (at least) a full correction to the incorrect program. (3) We extract the full code repairs, which we (4) pass to unit tests. The samples which pass or fail the unit tests allow us to evaluate (5) functional correctness and (6) closeness to the incorrect program.}
\label{fig:overview}
\end{figure}

In this section, we formalize the educational program repair task for large language models and present an evaluation procedure for that task (Figure \ref{fig:overview} shows an overview). We also introduce two educational scenarios of interest and propose two high-quality publicly available datasets that match these scenarios.

\subsection{Task and Evaluation Procedure}
\label{sec:task-eval}

\paragraph{Problem setup and motivations.} 
For this task, we consider a student working on a programming assignment. We have access to the problem description, associated unit tests, and a grader that employs these unit tests to provide summative feedback to students. When presented with an incorrect program that doesn't pass all unit tests, our goal is to utilize a language model to generate a repair to the student's code, addressing all the issues in the incorrect program. In essence, we view student program repair as a program synthesis problem, conditioned upon an existing (non-working) solution \cite{zhang2022repairing}.
Our primary objective is to ensure that the generated programs are not only functionally correct but also closely aligned with the original incorrect program. This closeness is akin to the preference for hint-generation systems to provide hints that are in proximity to the student's current state of knowledge \cite{gross2014example,rivers2017data,price2017evaluation}. The basic assumption is that enforcing this closeness will enhance the comprehensibility of the associated feedback \cite{clt_duran_2022}, such as hints generated by an Intelligent Tutoring System (ITS), or the natural language explanations generated by an LLM \cite{phung2023generating}.

\paragraph{Evaluation procedure.}
Generally, we consider each buggy program as an independent ``problem'', and use our language model to generate one or multiple repairs for each problem. Classically, (educational) program repair has reported performance in terms of the total number of buggy programs successfully repaired (i.e., the total number of problems solved) and the average edit distance between the buggy programs and their associated code repairs \cite{zhang2022repairing,phung2023generating}. In this work, we generate up to $n = 10$ different repairs and introduce two distinctions to the classical evaluation procedure. 

\subparagraph{pass@k}
First, following prior program synthesis research, we assess functional correctness using the pass@k estimator \cite{chen2021evaluating}. We first generate $n$ repairs. Then for a given $k$ ($k \leq n$), we generate all combinations of $k$ elements from the $n$ repairs, and count those with at least one repair passing all unit tests. 
Dividing this count by the total combinations gives the probability of a random sample of $k$ repairs fixing a buggy program. We average this probability for each program in our test set.

\subparagraph{rouge@k}
Second, recent research \cite{koutcheme_2023_evaluating} suggests that the Rouge score \cite{lin2004rouge} serves as a better repair quality indicator compared to sequence edit distance.  Building upon this insight, we introduce a novel evaluation metric, `rouge@k' mirroring `pass@k'. For each of the passing $n$ model generations, we calculate Rouge\footnote{Rouge embodies a family of metrics. For this study, we use the Rouge-L \protect\cite{lin2004rouge} score.} scores with the buggy program (using the buggy program as the reference `sentence'). Importantly, candidates that fail unit tests receive a score of 0. For some $k$, we generate all combinations of $k$ scores from the $n$ candidates and average the highest scores from each combination. 
We repeat this for each buggy program in our test set, and average the values to produce the final rouge@k score.

\subsection{Educational Scenarios and Datasets}
We propose two versions of the program repair task: a version where prior educational data is available, and one where no prior data is available. For both scenarios, we propose to rely on publicly available real-life datasets collected from various institutions. Due to space limitations, we refer the reader to the original dataset papers (resp. Appendix \ref{sec: app_dataset}) for curation (resp. processing) details.

\paragraph{Prior data available.}
In this scenario, we assume having access to prior data. Specifically, we define ``prior'' data as a repository of both correct and incorrect solutions submitted by students in preceding iterations of the same programming assignments, typically offered on a semester basis. Automated Program Repair has frequently relied on such prior data to construct repairs for students' buggy programs \cite{yang2019refactory,gulwani_automated_2018,yana_algo_corr,wang_data-driven_2017}. In our context, access to prior data offers significant advantages for large language models, whether for few-shot prompting \cite{zhang2022repairing} or supervised fine-tuning \cite{koutcheme_training_2023}. Additionally, it facilitates the proper division of our datasets into training, validation/development, and testing subsets. This mirrors a real-life scenario in which educational teams continually enhance their models each semester as they accumulate more data.
For this scenario, we propose harnessing the recently released FalconCode dataset \cite{falconcode}. Beyond its substantial scale, this dataset distinguishes itself through the presence of associated unit tests, and the inclusion of pertinent metadata (e.g. assignment difficulty level). Notably, it contains free-form assignments where students are not necessarily required to write functions. This change from the focus of prior work to repair functions \cite{koutcheme_automated_2023,zhang2022repairing} allows for a broader evaluation of LLM-based repair techniques. For evaluation purposes, we propose to utilize a smaller curated subset of 1,305 solutions for testing. It's worth noting that code benchmarks with fewer than 1000 problems are common in the field \cite{austin2021program,hendrycks2021measuring,lai2022ds1000,chen2021evaluating}, partly due to the computational and/or financial resources required for LLM-based evaluation techniques.

\paragraph{No prior data available.}
In the second version of the problem, we introduce a setup where models are tasked with fixing programs in a dataset where no prior data is available. This scenario aligns with contexts such as new courses (typically small) or courses where data about students' previous submissions is not retained.
To address this challenge, we propose leveraging a dataset collected at the National University of Singapore \cite{yang2019refactory}. This dataset comprises five assignments where students are required to write functions to solve specific tasks involving loops, conditions, and other standard introductory concepts. The reasons for selecting this dataset include an appropriate size, the selection of exercises with intermediate levels of difficulty, the presence of associated unit tests, and a good variety of issues in the submissions. We use a curated subset of 1238 incorrect submissions.

\section{Baseline and Experiments}
\label{sec: experiments}
In this section, we present a set of simple baseline models obtained from fine-tuning a variety of open-source pretrained language models of code. We leave the evaluation of instruction-tuned \cite{wei2022finetuned} and chat models \cite{ouyang2022training} for future work.

\subsection{Models}

We experiment with several decoder-only transformer models, using the CodeGen family of models \cite{nijkamp2023codegen} with 350M, and 2B parameters, as well as models from the StarCoder family \cite{li2023starcoder} with 164M, 1B, and 3B parameters. We chose these models for their relatively small size which allows them to fit fully (without quantization) inside a single consumer GPU.
Although we could evaluate the repair capabilities of these models using zero-shot and few-shot prompting \cite{joshi2022repair}, prior work suggests that their relatively small size (less than several billion parameters) prevents them from fully taking advantage of in-context examples \cite{brown2020language}. For this reason, we choose to finetune these models to repair student programs via supervised learning. 
Because most educational datasets do not contain ground truth repairs to students' incorrect programs, prior work in neural program repair \cite{gupta2019neural} suggested creating artificial repair examples by mapping each incorrect program in a dataset to the closest correct program submitted to the same assignment. Instead of using the closest correct program, we follow recent work \cite{koutcheme_training_2023} and map each incorrect program to the correction found by an Automated Repair Tool applied to the same dataset. Using an Automated Repair Tool (in our case, Refactory \cite{yang2019refactory}) allows for efficiently selecting a semantically close program while smoothing out many irrelevant syntactic differences (e.g., variable name conventions, needless reformulations) between the incorrect program and the chosen one." 

We trained our supervised baselines on the FalconCode dataset with the above-mentioned methodology, using the first two semesters of data as training and development sets (1299 and 1339 distinct programs respectively). Appendix \ref{sec: app_exp_details} details our exact training procedure including prompting strategy, and hyperparameter tuning. As an additional contribution, we released the code for conducting our experiments on GitHub\footnote{\href{https://github.com/KoutchemeCharles/gaied_nips23}{https://github.com/KoutchemeCharles/gaied\_nips23}}.

\subsection{Experiments and Results}

We report the performance of the supervised baselines on the test set of the FalconCode dataset. Following \cite{austin2021program}, we separate the results for two distinct assignment difficulty levels: ``easy'' and ``hard'' (with 767 and  538 incorrect programs respectively). We also report the performance of the models fine-tuned on FalconCode when prompted on the 1238 incorrect submissions of the Singapore dataset. The results for these three scenarios, including pass rates and Rouge scores, are presented in Table \ref{tab:results}. We can make the following observations.

\begin{table}[htbp]
    \caption{Functional and closeness results for our baselines models (higher is better).}
    \begin{subtable}[h]{\textwidth}
        \centering
        \caption{\textbf{Pass@k} for k = ${1, 5, 10}$}
        \resizebox{\columnwidth}{!}{
        \begin{tabular}{ll|rrr|rrr|rrr}
        \toprule
         &  & \multicolumn{3}{c}{falconcode\_easy} & \multicolumn{3}{c}{falconcode\_hard} & \multicolumn{3}{c}{singapore} \\
         model &  size & k = 1 & k = 5 & k = 10 & k = 1 & k = 5 & k = 10 & k = 1 & k = 5 & k = 10 \\
        \midrule
        starcoder & 164M & 6.88 & 12.46 & 14.68 & 20.27 & 35.93 & 41.85 & 2.54 & 7.62 & 10.46 \\
        codegen-mono & 350M & 12.08 & 23.19 & 27.88 & 13.68 & 28.28 & 33.77 & 3.72 & 12.92 & 18.90 \\
        starcoder & 1B & 16.91 & 30.38 & 36.06 & \textbf{26.44} & \textbf{44.94} & \textbf{50.72} & 7.35 & 21.02 & 28.77 \\
        codegen-mono & 2B & 16.10 & 25.99 & 30.67 & 13.85 & 25.92 & 30.38 & 11.24 & 24.97 & 31.63 \\
        starcoder & 3B & \textbf{19.11} & \textbf{31.68} & \textbf{37.73} & 14.63 & 29.88 & 36.51 & \textbf{12.29} & \textbf{29.37} & \textbf{36.14} \\
        \bottomrule
        \end{tabular}
        }
       \label{tab:week1}
    \end{subtable}
    \hfill
    \begin{subtable}[h]{\textwidth}
        \centering
        \caption{\textbf{Rouge@k} for k = ${1, 5, 10}$}
        \resizebox{\columnwidth}{!}{
        \begin{tabular}{ll|rrr|rrr|rrr}
        \toprule
         model &  size & k = 1 & k = 5 & k = 10 & k = 1 & k = 5 & k = 10 & k = 1 & k = 5 & k = 10 \\
        \midrule
        starcoder & 164M & 5.91 & 10.66 & 12.57 & 12.14 & 23.43 & 28.32 & 2.18 & 6.58 & 9.07 \\
        codegen-mono & 350M & 10.56 & 20.14 & 24.12 & 9.06 & 19.32 & 23.50 & 2.49 & 8.66 & 12.92 \\
        starcoder & 1B & 14.32 & 25.41 & 29.93 & \textbf{17.02} & \textbf{30.91} & \textbf{35.90} & 5.60 & 15.95 & 21.68 \\
        codegen-mono & 2B & 13.70 & 21.91 & 25.66 & 10.22 & 19.14 & 22.70 & 7.63 & 13.59 & 15.91 \\
        starcoder & 3B & \textbf{15.89} & \textbf{26.18} & \textbf{31.12} & 10.48 & 21.59 & 26.54 & \textbf{7.85} & \textbf{19.97} & \textbf{25.73} \\
        \bottomrule
        \end{tabular}
        }
        \label{tab:week2}
     \end{subtable}
     \label{tab:results}
\end{table}

\paragraph{Model performance scales with training compute and model sizes.} 
Within the same architectural family, performance scales with model sizes and number of generations. However, between different model families, the quality of the pretraining can make smaller models (e.g., starcoder 1B) outperform larger ones (e.g., codegen-mono 2B). The generalization capabilities of language models have been shown to scale smoothly with the amount of computing used during pretraining in proportion with model size and dataset size \cite{hoffmann2022training,touvron2023llama}.

\paragraph{Fine-tuned language models can overfit their datasets.}
Some models demonstrate superior performance on the `hard' section of the Falcon-code dataset compared to the `easy' section. We posit that this divergence stems from a training data imbalance, leading to `overfitting' on more challenging assignments \cite{tirumala2022memorization}. 
Notably, overfitting is more pronounced in smaller models, which may exhibit better performance than larger counterparts on the `hard' subset (e.g., codegen-mono 350M outperforming Codegen 2B on Falcon-code hard). However, this advantage does not extend to the smaller training subset (FalconCode easy) or to the Singapore dataset. 

\paragraph{Supervised training can transfer across datasets.} 
Despite being fine-tuned on the FalconCode dataset, our models managed to fix some of the buggy programs in another dataset featuring distinct problem types and other unique issues within the incorrect programs. Interestingly, beyond 1B parameters, the models reach similar performance on the FalconCode and Singapore datasets.

\section{Discussion and Conclusion}
We have introduced a benchmark for program repair using LLMs. However, the versatility of language models extends beyond program repair; they are also capable of generating natural language explanations paired with functioning solutions. Evaluating the quality of these explanations poses a challenge in the absence of ground truths or manual assessments. 
Although we cannot guarantee the accuracy of generated explanations with certainty, we assert that the relative quality of repairs can act as a reliable proxy for the relative quality of associated Natural Language Explanations (NLEs) when these NLEs are generated prior to the repairs. Previous research suggests that, for a given model, repairs of higher quality that are closely aligned with the student's code are more likely to be derived from in-context, high-quality explanations~\cite{wei2023chainofthought}. Conversely, generating natural language explanations from in-context, high-quality repairs~\cite{phung2023generating,koutcheme_2022_towards} -- first generating the repair and then generating explanations -- reduces the risk of introducing inaccuracies or inventing non-existent issues, thus mitigating the risk of hallucinations~\cite{bang2023multitask,hellas2023exploring}. Thus, while our primary focus is on program repair, our work also serves as a step towards improving the feedback that can be provided to students. 

While our baseline models provide valuable insights, it's essential to recognize the growing presence of powerful language models capable of program repair without additional training. In the source code modelling landscape, we observe a division between permissive (open-source) and non-permissive (proprietary) models. While prior work in educational program repair primarily focused on using non-permissive models \cite{zhang2022repairing,phung2023generating,phung2023generative,ahmed2022synfix,leinonen2023using}, we forecast a potential shift occurring with the emergence of robust open-source alternatives \cite{touvron2023llama,rozière2023code}. Historically, non-permissive models such as ChatGPT and OpenAI Codex have demonstrated state-of-the-art performance on benchmarks, but open-source alternatives are beginning to match or surpass them.
However, running such powerful permissive instruction-tuned and chat models requires significant computational resources (e.g. custom GPUs) due to their size. This is becoming less of a barrier due to the rise of open-source hosting services \cite{wolf2020huggingfaces}, the ability of smaller models to achieve high performance \cite{hoffmann2022training}, and the use of Parameter Efficient Finetuning Techniques (PEFT) \cite{houlsby2019parameterefficient}. While non-permissive models such as ChatGPT or OpenAI Codex offer high performance without the need for custom hardware, they also raise significant concerns about privacy and potential non-reliability due to model changes \cite{chen2023chatgpts}.

Prior studies in program repair have yielded valuable insights into the application of various methods.  However, meaningful comparisons between methods have been hindered by the lack of accessible, high-quality datasets. 
To overcome this limitation, we have curated the FalconCode and Singapore datasets specifically for program repair tasks relevant to educational settings.  Nevertheless, we acknowledge concerns that any individual educational benchmarking dataset can not fully capture the diversity of exercises and instructional approaches present in programming classes~\cite{vihavainen2014systematic,luxton2018introductory}. 
We feel it is important to emphasize that the purpose of these benchmarks should not be to establish a rigid hierarchy of method superiority but to serve as reference points for comparing methods in common scenarios, and for identifying issues in prior studies~\cite{sarsa2022empirical}. We believe that educational AI researchers can benefit from presenting and evaluating their novel techniques on both private datasets and the common benchmark, providing a well-rounded perspective.

Finally, we acknowledge several limitations to our current work.  First, both the FalconCode and Singapore datasets are limited to Python and represent data from only two institutions. Furthermore, syntax errors are currently not included due to our curation processes. We are also aware of the potential risk of contamination, a common challenge in code benchmarks \cite{golchin2023time}.  
While the FalconCode dataset is unlikely to be part of any model's pretraining dataset (as access requires a manual request via an online form), the Singapore dataset is readily available in a GitHub repository. Lastly, we recognize that assessing qualitative performance based solely on code closeness may introduce biases and may not always align with instructors' pedagogical objectives, such as teaching specific programming composition methods or emphasizing certain coding approaches (see e.g.~\cite{nelson2017comprehension,felleisen2018design,sorva2014based,edwards2020syntax}). 
To address these limitations, our future plans include introducing other high-quality datasets that cover a broad range of programming languages and scenarios (e.g. from online MOOC courses). We have also identified datasets suitable for evaluating both syntax and semantic errors, and we are working on annotating a subset of our datasets to provide accessible ground truth annotations \cite{koutcheme_2023_evaluating}. 
%
%
Looking further ahead, we intend to expand our benchmark to encompass a broader range of educational programming tasks \cite{phung2023generative} (e.g. hint generation), and we aim to establish an open online leaderboard \cite{wolf2020huggingfaces} for result tracking.
As educational program feedback presents distinctive challenges and real-life constraints, we anticipate cross-domain collaborations between educational program feedback research and pure NLP research to be essential.

\bibliographystyle{ACM-Reference-Format}  
\bibliography{biblio/pure_ai_biblio,biblio/eduai_biblio,biblio/cer_biblio}

\appendix
\section{Appendix}

\subsection{Dataset details}
\label{sec: app_dataset}
In this section, we detail the processing steps applied to each of our datasets.

\subsubsection{FalconCode}

We requested and acquired the FalconCode dataset from the authors' website \href{https://falconcode.dfcs-cloud.net/}{website}, which consists of multiple tables as detailed in the paper by \cite{falconcode}. The downloaded dataset encompasses 1,433,323 submissions written by 1,563 students across 478 distinct assignments spanning three semesters.

There are three levels of difficulty of assignments in the dataset: "skill", "lab" and "project". For this study, we excluded ``project'' assignments, which have high complexity, often require extensive code writing across multiple files, and lack automated tests. For all our analysis, we refer to ``skill'' and ``lab'' assignments as ``easy'' and ``hard'' (respectively). 
Several other assignments require access to external files (e.g, .csv files or .txt files) which the authors did not release in the current version of the dataset. However, we are currently in the process of retrieving this information. We excluded all assignments that require these external files.
We then removed students' intermediate runs (not submitted for grading). All other submissions present were automatically graded and assigned a score comprised between 0 and the exercise maximum score (typically 100). We removed all submissions which have a score of 0, as manual inspection reveals that these solutions are often the results of students' obvious mistakes (e.g. forgetting to print) or students' "trial and error". 
After removing intermediate runs and zero-passing codes, we have in total 270,478 submitted programs to 334 problems written by 1557 students.

Following \cite{yang2019refactory}, we selected only the final submissions for each student for each assignment. This resulted in our test set comprising submissions from students who did not complete the assignments successfully. While we acknowledge that this selection may not fully capture the range of difficulties students encounter during their attempts, it aligns with the idea that a student's last attempt often reflects their improved understanding of the problem.
Additionally, students are more likely to seek help after several unsuccessful attempts. Thus, our setup can be viewed as providing feedback to students as a last resort for elements they may not have grasped. In our dataset, this selection process yielded 118,764 distinct solutions. We recognize the potential for future work to explore more refined methods for automatically selecting meaningful solutions within large datasets.

For dataset splitting, we adopted a methodology similar to \cite{koutcheme_training_2023}. We divided the dataset by semester, designating the first two semesters (fall and spring 2021) for training and development purposes, respectively. The remaining semester constituted the testing set (spring 2022). Within each split, we independently removed submissions that shared identical abstract syntax tree (AST) structures after variable normalization. Table \ref{tab:dataset_statistics} provides an overview of the dataset statistics after this processing.

\begin{table}[htbp]
\caption{\textbf{FalconCode dataset statistics.}}
\centering
\begin{tabular}{l|ll|ll|ll}
subset                 & \multicolumn{2}{l|}{training} & \multicolumn{2}{l|}{development} & \multicolumn{2}{l}{testing} \\
difficulty             & easy          & hard          & easy           & hard            & easy         & hard         \\ \hline
\# correct solutons    & 4413          & 16463         & 4521           & 13552           & 3678         & 12164        \\
\# incorrect solutions & 398           & 901           & 622            & 717             & 538          & 767          \\
\# problems            & 89            & 104           & 87             & 68              & 83           & 89          
\end{tabular}
\label{tab:dataset_statistics}
\end{table}

Note that we intentionally also limited the size of our training and dev subsets to match the distribution of the test set. We encourage further experimentation with larger subsets of training datasets. 

\subsubsection{Singapore}

We obtained the Singapore dataset from the original paper's GitHub repository \cite{yang2019refactory}. The downloaded dataset initially contains 4225 submissions from 361 students. As with the FalconCode data, we removed similar submissions based on AST comparison on the Singapore dataset, which left us with a total of 2436 submissions. Table \ref{tab:sing_stat} displays the dataset statistics. It's important to note that the original dataset lacks information regarding which student submitted which assignment, and that we do not make use of the correct solutions.

\begin{table}[htbp]
\centering
\caption{\textbf{Singapore dataset statistics.}}
\begin{tabular}{@{}llllll|l@{}}
\toprule
Assignment              & 1   & 2   & 3   & 4   & 5   & Total\\ \midrule
\# correct solutions   & 480 & 157 & 152 & 186 & 223  & 1196 \\
\# incorrect solutions & 296 & 331 & 246 & 261 & 104 & 1238 \\ \bottomrule
\end{tabular}
\label{tab:sing_stat}
\end{table}

\subsection{Experiment details}
\label{sec: app_exp_details}
In this section, we detail how we created artificial code repair examples for supervised training (fine-tuning) of the used models, and how we trained and evaluated our models.

\subsubsection{Neural Automated Repair}
\label{sec:app-art}

Prior to training, we utilized the Refactory Automated Repair Tool (ART) \cite{yang2019refactory} to generate repairs for each incorrect program in the training and development datasets from FalconCode \cite{koutcheme_training_2023}. Detailed information about Refactory's functionality can be found in the original paper and its associated GitHub repository\footnote{\href{https://github.com/githubhuyang/refactory}{https://github.com/githubhuyang/refactory}}. 

We configured Refactory with online refactoring, sampling 100\% of correct student programs from both the training and development subsets, and enabled structure mutation and block repair phases (i.e., using the best settings available). Unlike in \cite{koutcheme_training_2023}, if ART couldn't successfully repair a faulty program, we followed the approach described in \cite{gupta2019neural}. In such cases, we employed the Rouge-L score \cite{koutcheme_2023_evaluating,lin2004rouge} to map the incorrect program to the nearest correct program. 

\subsubsection{Supervised training} 

We finetuned our models for next token prediction using HuggingFaces' transformers library \cite{wolf2020huggingfaces} using separate training and development sets. To ensure efficient training, we truncated each sequence to a maximum of 512 tokens. We decayed the learning rate with a cosine schedule, using an initial learning rate of 5e-5. We then performed a hyperparameter search, varying the number of epochs (1, 2, or 3) and batch sizes (4, 8, 16, or 32); the model selected for final evaluation with the testing set is the one showing the lowest validation loss on (a subset of) the development set. Our template prompt displayed in Figure \ref{fig:prompt} shows the structure of the prompts that we used to generate the repairs with the models. Notably, the last part of the prompt that includes the repair example found with ART is included only for the training and development set and excluded for testing sets.  It's important to note that the prompt used did not include the graders' feedback. In our future work, we will evaluate separately the effect of adding such feedback on program repair performance.

\begin{figure}[htbp!]
\centering
\begin{lstlisting}[basicstyle=\linespread{2},frame=single,escapeinside={(*@}{@*)}, ]
Repair the incorrect program. (*@\circled{1}@*)

{PROBLEM_DESCRIPTION} (*@\circled{2}@*)

**Incorrect program** (*@\circled{3}@*)

{BUGGY PROGRAM} (*@\circled{4}@*)

**Repaired code**: (*@\circled{5}@*)

{CORRECTED PROGRAM} (*@\circleddotted{6}@*)
\end{lstlisting}

\caption{The prompt template used for model training and evaluation: (1) a preamble orienting the model to the task of the program repair (although pretrained models are not instruction tuned, a preamble remains a useful signal), (2) the problem description extracted from the dataset, (3) an indication of the start of the buggy program, (4) the buggy program to fix, (5) an indication of the start of the repaired program, (6) the artificial repair found for the buggy program (see section \ref{sec:app-art}) -- included only in the training phase.}
\label{fig:prompt}
\end{figure}

\subsubsection{Evaluation}
We generated 10 repairs for each incorrect program in our test set using our prompt template shown in Figure \ref{fig:prompt}. The repaired programs were generated using top\_p nucleus sampling with a fixed temperature of 0.6, similarly as in \cite{chen2021evaluating} (all other parameters are default). We refer the reader to section \ref{sec:task-eval} for the rest of the evaluation procedure.

\subsubsection{Computational resources.} The training and evaluation pipelines were run individually and separately for each model on a single Nvidia Tesla A100 GPU using our institution research cluster.

\end{document}